\documentclass[11pt]{article}
\usepackage{latexsym}
\usepackage{amsfonts}
\usepackage{amsbsy}
\usepackage{mathrsfs}

\begin{document}

\title{Alternative symplectic structures for $SO(3,1)$ and $SO(4)$
four-dimensional BF theories}

\author{{\bf Merced Montesinos\thanks{Associate Member of the Abdus
Salam International Centre for Theoretical Physics, Trieste, Italy.}}\\
Departamento de F\'{\i}sica, Cinvestav, Av. Instituto Polit\'ecnico Nacional
2508,\\ San Pedro Zacatenco, 07360, Gustavo A. Madero, Ciudad de M\'exico,
M\'exico\\
merced@fis.cinvestav.mx}

\date{}

\maketitle

\begin{abstract}
The most general action, quadratic in the $B$ fields as well as in the
curvature $F$, having $SO(3,1)$ or $SO(4)$ as the internal gauge group for a
four-dimensional BF theory is presented and its symplectic geometry is
displayed. It is shown that the space of solutions to the equations of motion
for the BF theory can be endowed with symplectic structures alternative to the
usual one. The analysis also includes topological terms and cosmological
constant. The implications of this fact for gravity are briefly discussed.
\end{abstract}

\section{Introduction}\label{I}
The canonical analysis of a given classical theory is the first step towards
its canonical quantization and so it is worthwhile to perform it. In this
paper, it is adopted the point of view that what defines a dynamical system is
its equations of motion, which do {\it not} uniquely fix the symplectic
structure on the phase space of the dynamical system under consideration
\cite{morandi,monger}. This simple fact can be clearly appreciated even in
systems with a finite number of degrees of freedom. Take, for instance, the
equations of motion for the two-dimensional isotropic harmonic oscillator
\begin{eqnarray}\label{emho}
{\dot x} = \frac{p_x}{m}\, , \quad {\dot y} = \frac{p_y}{m}\, \quad {\dot p}_x
= - m \varpi^2 x\, , \quad {\dot p}_y = - m \varpi^2 y\, ,
\end{eqnarray}
where $m\neq 0$ is the mass of the particle, $\varpi$ the angular frequency
and the dot `{$ \cdot $}' stands for total derivative with respect to the
Newtonian time $t$. The phase space for the system is $\Gamma=\mathbb{R}^4$.
The usual Hamiltonian formulation for equations (\ref{emho}) comes from the
action principle
\begin{eqnarray}\label{ho1}
S [x,y,p_x,p_y] = \int^{t_2}_{t_1} d t \left [ {\dot x} p_x + {\dot y} p_y - H
\right ]\, ,
\end{eqnarray}
where $H$ is the energy for the system. However, the equations of motion
(\ref{emho}) can alternatively also be obtained from the action principle
\begin{eqnarray}\label{ho2}
S_1 [x,y,p_x,p_y] &=&  \int^{t_2}_{t_1}  dt \left [ {\dot x} p_y + {\dot y}
p_x -
H_1 \right ]\, , \nonumber\\
H_1 &=& \frac{p_x p_y}{m} + m \varpi^2 xy \, .
\end{eqnarray}
This simple example allows it to emphasize some points:
\begin{enumerate}
\item
both actions (\ref{ho1}) and (\ref{ho2}) are functionals of the {\it same}
variables: $x$, $y$, $p_x$, and $p_y$,
\item
both actions (\ref{ho1}) and (\ref{ho2}) yield the same equations of motion
(\ref{emho}),
\item
in spite of the fact that actions (\ref{ho1}) and (\ref{ho2}) yield the same
equations of motion, they both define two, {\it different}, symplectic
structures on $\Gamma=\mathbb{R}^4$. In fact, the symplectic structure defined
by action (\ref{ho1}) is $\omega= d \theta = d (p_x d x + p_y dy ) = dp_x
\wedge dx + d p_y \wedge d y$ while the one defined by action (\ref{ho2}) is
$\omega_1= d \theta_1 = d (p_y dx + p_x d y ) = d p_y \wedge dx + d p_x \wedge
d y$. It is clear that $\omega \neq \omega_1$. Moreover, suppose that
$\varphi: \mathbb{R}^4 \rightarrow \mathbb{R}^4$ is a map such that
$\varphi^{\ast} \omega = \omega_1$. This fact together with $\omega_1 \neq
\omega$ implies that $\varphi^{\ast} \omega \neq \omega$ which means that
$\varphi$ is not a symplectomorphism.
\item
it is also important to mention that we are {\it not} making a redefinition of
variables, i.e., the $x$ that appears in $\omega$ is the same $x$ that appears
in $\omega_1$ and so on.
\end{enumerate}
In summary, the example exhibits the fact that different Hamiltonian
formulations can be given to the same set of equations of motion and that one
way to do this is to use an action principle. Thus, the action principle plays
a double role: on one hand it gives us the equations of motion and on the
other it also fixes a particular symplectic geometry on the phase space, which
is usually not spelled out, and most of the times underestimated. In our
opinion, this fact is not just an academic one because the knowledge of the
symplectic structure is the first step towards the canonical quantization of
the theory (for more details on the quantum theories of the two-dimensional
isotropic harmonic oscillator, see Refs. \cite{morandi,monger}). The fact that
a given set of equations of motion admits various Hamiltonian formulations has
also been studied for generally covariant systems with a finite number of
degrees of freedom \cite{mo,solo} and a proposal for the Hamilton-Jacobi
theory of dynamical systems having non-canonical symplectic structures has
also been made \cite{aldo,aldo2}.

So, it is natural to ask if the space of solutions of gauge field theories
might be endowed with various symplectic structures, and we think that
four-dimensional BF theories are very good models to explore this idea.
Four-dimensional BF theories having $SO(3,1)$ or $SO(4)$ as the internal
symmetry group are relevant by themselves and also by their close relationship
with general relativity \cite{pleban,Mike,Pietri,Capo}. This theory is defined
by the equations of motion
\begin{eqnarray}\label{emotion}
F^{IJ}=0 \, , \quad D B^{IJ}=0 \, .
\end{eqnarray}
Here, $I,J,K,\ldots=0,1,2,3$ are Lorentzian (or Euclidean) indices which are
raised and lowered with the internal metric $\eta_{IJ}$,
$(\eta_{IJ})=\mbox{diag} (\sigma,1,1,1)$ with $\sigma=-1$ for Lorentzian and
$\sigma=1$ for Euclidean signatures, $B^{IJ}=\frac12 B_{\alpha\beta}\,^{IJ} d
x^{\alpha} \wedge d x^{\beta}$ is a set of six 2-forms on account of the
antisymmetry of $B^{IJ}$ in the internal indices, $B^{IJ}=-B^{JI}$, the
indices $\alpha, \beta, \ldots$ are spacetime indices and the coordinates
$x^{\alpha}$ label the points $x$ of the 4-dimensional manifold $\mathscr{M}$,
$F^I\,_J [A] = d A^I\,_J + A^I\,_K \wedge A^K\,_J$ is the curvature of the
connection 1-forms $A^I\,_J$, $F^I\,_J = \frac12 F_{\alpha\beta}\,^I\,_J d
x^{\alpha} \wedge d x^{\beta}$, and $D B^{IJ}= d B^{IJ} + A^I\,_K \wedge
B^{KJ} + A^J\,_K \wedge B^{IK}$ (see the appendix).

The equations of motion (\ref{emotion}) are usually obtained from the action
\cite{horo} (see also \cite{rovbook})
\begin{eqnarray}\label{BFaction}
S[A,B] & = & a_1  \int_{\mathscr{M}} B^{IJ} \wedge F_{IJ} [A],
\end{eqnarray}
(for an analysis of the symmetries of the action (\ref{BFaction}) see Ref.
\cite{symmetry})). Let ${\bf \Gamma_{\mbox{cov}}}$ be the space of solutions
to the equations of motion (\ref{emotion}) on ${\mathscr M}$ (for details on
the notation and conventions used in the covariant canonical formalism see
Refs. \cite{Crn,Lee,bombelli}). Let $({\overline \delta} A, {\overline \delta}
B )$ be tangent vectors to the space of histories ${\bf \mathscr F}$ (formed
by all smooth configurations which do not necessarily satisfy the equations of
motion). Assuming that the equations of motion (\ref{emotion}) hold, then the
first order change of the Lagrangian 4-form $L=a_1 B^{IJ} \wedge F_{IJ}[A]$ is
\cite{preprint}
\begin{eqnarray}
{\overline \delta} L \mid_{{\bf {\Gamma}_{\mbox{cov}}}} = d \left ( a_1 B^{IJ}
\wedge {\overline \delta} A_{IJ} \right ).
\end{eqnarray}
This allows it to define a 1-form ${\bf \Theta}$ on the space of histories
${\bf \mathscr F}$ via
\begin{eqnarray}
{\bf \Theta} \left ( {\overline \delta} \right ) := \int_{\Sigma} a_1 B^{IJ}
\wedge {\overline \delta} A_{IJ}.
\end{eqnarray}
The symplectic structure ${\bf \Omega}$ is the pullback to ${\bf
\Gamma_{\mbox{cov}}}$ of the curl  of ${\bf \Theta}$ on the space of
histories, and it is given by \cite{baez,preprint}
\begin{eqnarray}\label{sym1}
{\bf \Omega} &=& a_1 \int_{\Sigma} \left ( \delta_1 B^{IJ} \wedge \delta_2
A_{IJ} - \delta_2 B^{IJ} \wedge \delta_1 A_{IJ} \right ) \nonumber\\
&=& 2 a_1 \int_{\Sigma} \left ( \delta_1 B_{0i} \wedge \delta_2 A^{0i} +
\delta_1 B_i \wedge \delta_2 \Gamma^i \right ) - \delta_1 \longleftrightarrow
\delta_2\, ,
\end{eqnarray}
for all tangent vectors $\delta_1$ and $\delta_2$ to ${\bf
\Gamma_{\mbox{cov}}}$, and the manifold $\mathscr{M}$ has been assumed to have
the topology $\mathscr{M}=\Sigma \times \mathbb{R}$ where $\Sigma$ stands for
`space' and ${\mathbb R}$ for `time'. Here, $B^i := -\frac12
\varepsilon^i\,_{jk} B^{jk}$ and $\Gamma^i := -\frac12 \varepsilon^i\,_{jk}
A^{jk}$. On the other hand, Dirac canonical analysis of the action
(\ref{BFaction}) implies that the theory has two sets of first class
constraints, ${\widetilde \Psi}^{IJ}:= D_a {\widetilde \Pi}^{aIJ} \approx 0$
and ${\widetilde \Psi}^a\,_{IJ}= \frac12 \eta^{abc} F_{bcIJ} (A) \approx 0$
where $(A_{aIJ}, {\widetilde \Pi}^{bKL})$ are canonically conjugate variables.
The constraints ${\widetilde \Psi}^a\,_{IJ}$ are reducible because of
${\widetilde \Phi}_{IJ} := D_a {\widetilde \Psi}^{aIJ} = 0$. So, taking the
reducibility of the constraints ${\widetilde \Psi}^a\,_{IJ}$ into account, the
counting of degrees of freedom of the theory yields zero degrees of freedom
per point of $\Sigma$ \cite{caicedo} (see also \cite{preprint}). Thus, the
field theory defined by the action of Eq. (\ref{BFaction}) has no local
degrees of freedom, the theory is sensitive only to the global degrees of
freedom associated with non-trivial topologies of the manifold $\mathscr{M}$
itself and topologies of the gauge bundle.

The BF theory defined by action (\ref{BFaction}) involves, besides the
$B^{IJ}$ and $F_{IJ}[A]$ fields, the Killing-Cartan metric $\eta_{IJKL}$ of
the Lie algebra $so(3,1)$ or $so(4)$. In this paper, the most general action
principle for the four-dimensional BF theory which is quadratic in $B^{IJ}$ as
well as in $F_{IJ}[A]$ involving {\it both} the Killing-Cartan metric
$\eta_{IJKL}$ and the volume form $\varepsilon_{IJKL}$ is constructed and its
symplectic geometry is displayed.

\section{$SO(3,1)$ and $SO(4)$ four-dimensional BF theories}\label{II}
Besides the action principle (\ref{BFaction}) usually employed to define the
BF theory, it is also possible to define the BF theory using the action
\cite{solo}
\begin{eqnarray}\label{a2}
S_1 [A,B] =  a_2 \int_{\mathscr{M}} {\ast B}^{IJ} \wedge F_{IJ}[A] \, ,
\end{eqnarray}
where ${\ast B}^{IJ}= \frac12 \varepsilon^{IJ}\,_{KL} B^{KL}$ is the dual of
$B^{IJ}$. Here $\varepsilon_{IJKL}$ is the volume 4-form associated with the
metric $\eta_{IJ}$, $\varepsilon_{0123}=\epsilon$ and
$\varepsilon^{0123}=\sigma \epsilon$ (see the appendix). The equations of
motion obtained from the action (\ref{a2}) are
\begin{eqnarray}
\ast F_{IJ} = 0 \, , \quad D \ast B^{IJ} = 0 \, ,
\end{eqnarray}
which after the application of the star operation ``$\ast$"  reduce to eqs.
(\ref{emotion}). Therefore, the theories defined by (\ref{BFaction}) and
(\ref{a2}) have the same space of solutions ${\bf \Gamma_{\mbox{cov}}}$ to the
equations of motion. What about the geometry? The symplectic potential on the
space of histories ${\bf \mathscr F}$ is now
\begin{eqnarray}
{\bf \Theta}_1 ({\overline \delta}) = \int_{\Sigma} a_2 \ast B^{IJ} \wedge
{\overline \delta} A_{IJ},
\end{eqnarray}
and the symplectic structure on the space of solutions ${\bf
\Gamma_{\mbox{cov}}}$ is given by
\begin{eqnarray}\label{sym2}
{\bf \Omega}_1 &=& a_2 \int_{\Sigma} \left ( \delta_1 \ast B^{IJ} \wedge
\delta_2 A_{IJ} - \delta_2 \ast B^{IJ} \wedge \delta_1 A_{IJ}\right )
\nonumber\\
&=& 2 a_2 \epsilon \int_{\Sigma} \left ( - \delta_1 B_i \wedge \delta_2 A^{0i}
- \sigma \delta_1 B_{0i} \wedge \delta_2 \Gamma^i \right ) - \delta_1
\longleftrightarrow \delta_2 \, ,
\end{eqnarray}
for all tangent variations $\delta_1$ and $\delta_2$ to ${\bf
\Gamma_{\mbox{cov}}}$ (cf equation (\ref{sym1})). From equation ${\bf
\Theta}_1$ it is clear that the momenta canonically conjugate to $A_{IJ}$ are
different from those of the action (\ref{BFaction}). Moreover, the symplectic
structure of equation (\ref{sym2}) is different from the symplectic structure
of Eq. (\ref{sym1}). This is just a reflection of the fact that the space of
solutions ${\bf \Gamma_{\mbox{cov}}}$ of the equations of motion of a
dynamical system can be endowed with more than one symplectic structure
\cite{monger,mo,solo}. As it was stressed in Section \ref{I} and in Ref.
\cite{preprint}, one way to fix the symplectic structure is to choose a
particular action principle.

Moreover, it is also possible to take a linear combination of both actions
(\ref{BFaction}) and (\ref{a2}) which gives rise to the action principle
\begin{eqnarray}\label{a3}
S_2 [A,B] & = & a_1 \int_{\mathscr{M}} B^{IJ} \wedge F_{IJ} [A] + a_2
\int_{\mathscr{M}} {\ast B}^{IJ} \wedge F_{IJ} [A] \, .
\end{eqnarray}
The field variables which are taken as independent variables in action
(\ref{a3}) depend on the relationship between the parameters $a_1$ and $a_2$.
This is discussed in what follows:

i) {\it actions involving self-dual or anti-self-dual variables}. By
decomposing $B^{IJ}$ and $A^{IJ}$ in terms of its self-dual $^{+}B^{IJ}$ and
anti-self-dual $^{-}B^{IJ}$ parts, $^{+}B^{IJ}= \frac12 \left ( B^{IJ} - i
{\ast B}^{IJ} \right )$ and $^{-}B^{IJ}=\frac12 \left ( B^{IJ} + i {\ast
B}^{IJ} \right )$ in the case of Lorentzian signature $\sigma=-1$, the action
(\ref{a3}) acquires the form
\begin{eqnarray}
S [A,B] =  ( a_1 + i a_2 ) \int_{\mathscr{M}} {^{+} B^{IJ}} \wedge F_{IJ}
[{^+A}] + ( a_1 - i a_2 ) \int_{\mathscr{M}} {^{-} B^{IJ}} \wedge F_{IJ}
[{^-A}]\, .
\end{eqnarray}
Self-dual variables. If $i a_2 = a_1$, last action reduces to
\begin{eqnarray}
S [ ^{+}A, ^{+}B ] & = & 2 a_1 \int_{\mathscr{M}}  {^{+}B^{IJ}} \wedge F_{IJ}
[^{+}A] \, ,
\end{eqnarray}
where $F^I\,_J [^{+}A] = d ^{+}A^I\,_J + ^{+}A^I\,_K \wedge ^{+}A^K\,_J$ is
the curvature of the self-dual connection 1-form $^{+}A^I\,_J$.\\
Anti-self-dual variables. If $i a_2 = - a_1$, the action reduces to
\begin{eqnarray}
S [ ^{-}A, ^{-}B ] & = & 2 a_1 \int_{\mathscr{M}}  {^{-}B^{IJ}} \wedge F_{IJ}
[^{-}A] \, ,
\end{eqnarray}
where $F^I\,_J [^{-}A] = d ^{-}A^I\,_J + ^{-}A^I\,_K \wedge ^{-}A^K\,_J$ is
the curvature of the self-dual connection 1-form $^{-}A^I\,_J$.

ii) {\it action involving real variables}. It is, however, still possible to
rely on the Minkowskian signature $(-,+,+,+)$ using real variables, keeping
$a_2/a_1 =: 1/\gamma$ real, and therefore with values distinct to the
exceptional values $\pm i$, $\gamma \neq \pm i$ for the Lorentzian signature
$\sigma=-1$ ($\gamma\neq \pm 1$ for the Euclidean signature $\sigma=1$). In
this case, the equations of motion
\begin{eqnarray}
{\mathscr E}_{IJ} [A] := a_1 F_{IJ} [A] + \frac{a_1}{\gamma} {\ast F}_{IJ} [A]
=
0 \, , \nonumber\\
{\cal E}^{IJ} [A,B] := a_1 D B^{IJ} + \frac{a_1}{\gamma} D {\ast B}^{IJ} = 0
\, ,
 \end{eqnarray}
obtained from action (\ref{a3}), reduce, on account of $\gamma\neq \pm i$, to
$F_{IJ} [A]=0$ and $D B^{IJ}=0$, which are equations (\ref{emotion}).
Therefore, the field theories defined by actions (\ref{BFaction}) and
(\ref{a3}) have the same space of solutions ${\bf \Gamma_{\mbox{cov}}}$ of the
equations of motion, i.e., ${\bf \Gamma_{\mbox{cov}}}$ is independent of
$\gamma$. However, the symplectic structure ${\bf \Omega}_2$ does depend on
$\gamma$. In fact, the symplectic potential on the space of histories is now
\begin{eqnarray}
{\bf \Theta}_2 ({\overline \delta}) = \int_{\Sigma} a_1 \left ( B^{IJ} +
\frac{1}{\gamma} \ast B^{IJ} \right ) \wedge {\overline \delta} A_{IJ},
\end{eqnarray}
while the symplectic structure on ${\bf \Gamma_{\mbox{cov}}}$ is
\begin{eqnarray}\label{sym3}
{\bf \Omega}_2 &=& a_1 \int_{\Sigma} \left [ \delta_1 \left ( B^{IJ} +
\frac{1}{\gamma} \ast B^{IJ} \right ) \wedge \delta_2 A_{IJ} - \delta_2 \left
( B^{IJ} + \frac{1}{\gamma} \ast B^{IJ} \right ) \wedge
\delta_1 A_{IJ} \right ] \nonumber\\
&=& 2 a_1 \int_{\Sigma} \left [ \delta_1 \left ( B_{0i} -
\frac{\epsilon}{\gamma} B_i \right ) \wedge \delta_2 A^{0i} + \delta_1 \left (
B_i - \frac{\epsilon \sigma}{\gamma} B_{0i} \right ) \wedge \delta_2 \Gamma^i
\right] - \delta_1 \longleftrightarrow \delta_2 \, ,
\end{eqnarray}
for all tangent variations $\delta_1$ and $\delta_2$ to ${\bf
\Gamma_{\mbox{cov}}}$ (cf equations (\ref{sym1}) and (\ref{sym2})).

We end this section by rewriting the action of equation (\ref{a3}) in the form
\begin{eqnarray}\label{rdp}
S_2 [A,B] & = & a_1 \int_{\mathscr{M}} s_{IJKL} B^{IJ} \wedge F^{KL} [A] \, ,
\end{eqnarray}
where
\begin{eqnarray}
s_{IJKL} := \frac12 \left ( k_{IJKL} + \frac{a_2}{a_1} \, \varepsilon_{IJKL}
\right )\, ,
\end{eqnarray}
is a metric on the Lie algebra $so(3,1)$. Here, $k_{IJKL}:= k \left ( X_{IJ} ,
X_{KL} \right ) = \eta_{IK} \eta_{JL} - \eta_{JK} \eta_{IL}$ is the
Killing-Cartan metric on $so(3,1)$ and $\varepsilon \left ( X_{IJ} , X_{KL}
\right ) := \varepsilon_{IJKL}$ is also a metric on $so(3,1)$ induced by the
volume 4-form, i.e., as a vector space $so(3,1)$ admits two, different, metric
structures. The infinitesimal generators $X_{IJ}=-X_{JI}$ of $so(3,1)$ satisfy
$[ X_{IJ} , X_{KL} ] = \eta_{JK} X_{IL} - \eta_{IK} X_{JL} - \eta_{JL} X_{IK}
+ \eta_{IL} X_{JK}$ (see the appendix). The same holds for $so(4)$, of course.

In summary, all three actions (\ref{BFaction}), (\ref{a2}), and (\ref{a3})
give rise to the same equations of motion (\ref{emotion}) and therefore they
all have the same space of solutions ${\bf \Gamma_{\mbox{cov}}}$. In this
sense they define the same dynamical system. However, the symplectic
structures that come from these actions are different from each other simply
because the three symplectic structures (\ref{sym1}), (\ref{sym2}), and
(\ref{sym3}) have different expressions when they are expressed with respect
to the same coordinates $(A_{IJ},B^{KL})$ which label the points of the space
of solutions ${\bf \Gamma_{\mbox{cov}}}$, i.e., ${\bf \Omega} \neq {\bf
\Omega}_2 \neq {\bf \Omega}_1$ (this is the analogue situation of $\omega$ and
$\omega_1$ for the two-dimensional isotropic harmonic oscillator mentioned in
section \ref{I}). Note that if a redefinition of the fields were done, for
instance, in the action (\ref{a2}), ${B'}^{IJ} := \varepsilon_{IJKL} B^{KL}$,
one would be tempted to say that action (\ref{a2}) written in the variables
${B'}$ and $A$, $\int_{\mathscr M} {B'}^{IJ} \wedge F_{IJ}[A]$, `is the same
as action (\ref{BFaction}).' This conclusion is, however, not correct. Of
course, it is possible to make that change of variables, but that is not what
we are doing in this paper and also that redefinition is not relevant for the
present analysis. We emphasize again that all three actions (\ref{BFaction}),
(\ref{a2}), and (\ref{a3}) are functionals of the {\it same} variables
$B^{IJ}$ and $A^I\,_J$ but the functional dependency of these actions on the
fields $B^{IJ}$ and $A^I\,_J$ is {\it not} the same which translates in having
{\it different} symplectic structures on the {\it same} space of solutions
${\bf \Gamma_{\mbox{cov}}}$ (actions (\ref{BFaction}), (\ref{a2}), and
(\ref{a3}) are analogues of actions (\ref{ho1}) and (\ref{ho2}) for the
two-dimensional isotropic harmonic oscillator).

It is interesting to note that for the action $S_2[A,B]$, the $\gamma$
dependency in both the symplectic potential ${\bf \Theta}_2$ and the
symplectic structure ${\bf \Omega}_2$ can also be absorbed in the connection
1-form
\begin{eqnarray}
{\bf \Theta}_2 ({\overline \delta}) &=& \int_{\Sigma} a_1 B^{IJ} \wedge
{\overline \delta} \left ( A_{IJ} +
\frac{1}{\gamma} {\ast A}_{IJ} \right ), \nonumber\\
{\bf \Omega}_2 &=& \int_{\Sigma} \left [ a_1 \delta_1 B^{IJ} \wedge \delta_2
\left ( A_{IJ} + \frac{1}{\gamma} {\ast A}_{IJ} \right ) - a_1 \delta_2 B^{IJ}
\wedge \delta_1 \left ( A_{IJ} + \frac{1}{\gamma} {\ast A}_{IJ} \right )
\right ].
\end{eqnarray}
This is just a reflection of Darboux's theorem. Due to the fact that ${\bf
\Omega} \neq {\bf \Omega}_2 \neq {\bf \Omega}_1$ one should expect quantum
theories unitarily inequivalent even though one has the same classical
dynamics. Whether or not the $\gamma$ parameter plays a role in the spin foam
formalism \cite{rovbook} similar to the one that the Barbero-Immirzi parameter
plays in loop quantum gravity is an open question as per as author's knowledge
\footnote{Note, incidentally, that in a three-dimensional spacetime ${\cal M}$
which is locally homogeneous (with curvature proportional to the cosmological
constant $\lambda$) there are {\it two} Chern-Simons Lagrangians which yield
the same classical equations of motion. The quantum theories, on the other
hand, are different from each other \cite{witten}. These two Chern-Simons
Lagrangians (and their linear combination) resemble the BF actions
(\ref{BFaction}), (\ref{a2}), and (\ref{a3}).}.

\section{Adding topological terms}\label{III}
Due to the fact the BF theory has no local degrees of freedom, the non trivial
topologies of the manifold ${\mathscr M}$ itself and the topologies of the
gauge bundle might be relevant. These two aspects can be `detected' by adding
to the Lagrangian of action (\ref{a3}) the second Chern class $c_2 [A] :=
\frac{1}{8\pi^2} F^I\,_J [A] \wedge F^J\,_I [A]$ (which `sees' the topology of
the gauge bundle) as well as with the Euler class $e [A] := \frac{1}{32 \pi^2}
\varepsilon_{IJKL} F^{IJ} [A] \wedge F^{KL} [A]$ (which `sees' the topology of
${\cal M}$). Even though the inclusion of these two terms is so obvious, a
systematic Hamiltonian analysis of the action
\begin{eqnarray}\label{a4}
S_3 [A,B] &=&  a_1 \int_{\mathscr{M}} B^{IJ} \wedge F_{IJ} [A] + a_2
\int_{\mathscr{M}} {\ast B}^{IJ} \wedge F_{IJ}[A] \nonumber\\
&& \mbox{} + \frac{\theta_1}{8\pi^2} \int_{\mathscr{M}} F^I\,_J [A] \wedge
F^J\,_I [A] + \frac{\theta_2}{16 \pi^2} \int_{\mathscr{M}} {\ast F}^{IJ}[A]
\wedge F_{IJ} [A],
\end{eqnarray}
does not exist in literature (see, however, Ref. \cite{preprint}), as per as
author's knowledge\footnote{If the spacetime ${\mathscr M}$ does have a
boundary $\partial {\mathscr M}$ then the action acquires a contribution from
$\partial {\mathscr M}$. In this case the action is $S_3 [A,B] +
\frac{\theta_1}{16 \pi^2} \int_{\partial {\mathscr M} } \eta_{IJKL}
\omega^{IJ} \wedge f^{KL}[\omega]$ - $\frac{\theta_2}{32 \pi^2} \int_{\partial
\mathscr M} \varepsilon_{IJKL} \omega^{IJ} \wedge f^{KL} [\omega]$ with
$f^I\,_J [ \omega ] = d \omega^I\,_J + \frac{2}{3} \omega^I\,_K \wedge
\omega^K\,_J$. Thus, there are {\it two} Chern-Simons Lagrangians. The first
term in the boundary Lagrangian which involves the Killing-Cartan metric
$\eta_{IJKL}$ is the usual Chern-Simons Lagrangian. The second term in the
boundary Lagrangian which involves the other metric $\varepsilon_{IJKL}$ might
also be considered of the Chern-Simons type.}.

i) {\it action involving real variables}. Once again, the field theory defined
by the action (\ref{a4}) can involve self-dual, anti-self-dual or real
variables depending on the relationship among the values of the parameters
involved. This case will be discussed few lines below. For the time being, let
us assume that the parameters $a_1$, $a_2$, $\theta_1$, and $\theta_2$ are
such that one is dealing with real variables $A^I\,_J$ and $B^{IJ}$ and that
all of them are independent at the level of the action principle (\ref{a4}).
Under this assumption, the equations of motion obtained from the action
(\ref{a4}) reduce to Eqs. (\ref{emotion}) on account of the Bianchi identities
$D F^{IJ}=0$. Therefore, the space of solutions ${\bf \Gamma_{\mbox{cov}}}$ is
the same one as before. Nevertheless, the symplectic potential ${\bf
\Theta}_3$ on the space of histories ${\bf \mathscr F}$ is now given by
\begin{eqnarray}\label{4pot}
{\bf \Theta}_3 ({\overline \delta}) = \int_{\Sigma} \left ( a_1 B^{IJ} +
\frac{a_1}{\gamma} {\ast B}^{IJ} - \frac{\theta_1}{4\pi^2} F^{IJ} +
\frac{\theta_2}{8 \pi^2} {\ast F}^{IJ} \right ) \wedge {\overline \delta}
A_{IJ},
\end{eqnarray}
which reduces to
\begin{eqnarray}
{\bf \Theta}_3 (\delta) = \int_{\Sigma} \left ( a_1 B^{IJ} + a_2 {\ast B}^{IJ}
\right ) \wedge \delta A_{IJ},
\end{eqnarray}
for tangent variations $\delta$ to the space of solutions ${\bf
\Gamma_{\mbox{cov}}}$. Even though the momenta canonically conjugate to
$A^I\,_J$ in the space of histories are modified, the symplectic structure
${\bf \Omega}_3$ computed by taking the pullback to ${\bf
\Gamma_{\mbox{cov}}}$ of the curl of ${\bf \Theta}_3$ on the space of
histories ${\bf \mathscr F}$ is exactly the same found in the preceding
section, namely, ${\bf \Omega_2}$. So, neither the space of solutions ${\bf
\Gamma_{\mbox{cov}}}$ nor the symplectic structure on it is sensitive to the
parameters $\theta_1$ and $\theta_2$. As is clear from (\ref{4pot}) the
momenta canonically conjugate to $A^I\,_J$ do depend on $\theta_1$ and
$\theta_2$.

However, there is a set of canonical coordinates which is peculiar in the
sense that it combines the various parameters involved in action (\ref{a4}).
The algebraic reason for this fact can be clearly appreciated by rewriting
action (\ref{a4}) in the form
\begin{eqnarray}
S [A,B]  = a_1 \int_{\mathscr{M}} s_{IJKL} B^{IJ} \wedge F^{KL} [A] -
\frac{\theta_1}{8 \pi^2} \int_{\mathscr{M}} g_{IJKL} F^{IJ} [A] \wedge F^{KL}
[A] \, ,
\end{eqnarray}
where
\begin{eqnarray}
g_{IJKL} := \frac12 \left ( k_{IJKL} - \frac{\theta_2}{2 \theta_1}
\varepsilon_{IJKL} \right ),
\end{eqnarray}
is a metric on the Lie algebra $so(3,1)$ or $so(4)$. {\it A priori}, there is
no reason to decide if to have one single metric on the Lie algebra $so(3,1)$
or $so(4)$ is more natural than having two metrics on it. A democratic
criterium would imply to have a single metric only, $s_{IJKL}=g_{IJKL}$, which
is relevant because then the parameters are related among themselves
\begin{eqnarray}\label{rela}
\frac{1}{\gamma}:= \frac{a_2}{a_1} = - \frac{\theta_2}{2 \theta_1},
\end{eqnarray}
and so, $a_2/a_1 = 1/\gamma$ is sensitive, through $\frac{a_2}{a_1}=
\frac{1}{\gamma} = - \frac{\theta_2}{2 \theta_1}$, to the topological aspects
of the manifold $\mathscr{M}$ itself and the global aspects of the gauge
bundle. This case also allows it to define a new set of canonical coordinates
because the symplectic potential ${\bf \Theta}_3 ({\overline \delta})$ in the
space of histories ${\bf \mathscr F}$ acquires the form
\begin{eqnarray}\label{canc}
{\bf \Theta}_3 ({\overline \delta}) = \int_{\Sigma} \left ( a_1 B^{IJ} -
\frac{\theta_1}{4 \pi^2} F^{IJ} \right ) \wedge {\overline \delta } \left (
A_{IJ} + \frac{1}{\gamma} {\ast A}_{IJ} \right ).
\end{eqnarray}

Of course, the relationship between the parameters (\ref{rela}) disappears as
well as the possibility of choosing the canonical coordinates (\ref{canc}) if
the two metrics on the Lie algebra of $so(3,1)$ or $so(4)$ are distinct from
each other, i.e., $s_{IJKL} \neq g_{IJKL}$.

ii) {\it actions involving self-dual and anti-self-dual variables}. By
decomposing the fields $B^{IJ}$ and $A_{IJ}$ into their self-dual and
anti-self-dual parts, the action (\ref{a4}) acquires the form (restricting the
analysis to the Lorentzian signature $\sigma=-1$)
\begin{eqnarray}\label{previa}
S[A,B] &=& ( a_1 + i a_2 ) \int_{\mathscr{M}} {^{+} B^{IJ}} \wedge F_{IJ}
[{^+A}] + ( a_1 - i a_2 ) \int_{\mathscr{M}} {^{-}
B^{IJ}} \wedge F_{IJ} [{^-A}] \nonumber\\
&& \mbox{} - \left ( \frac{\theta_1}{8\pi^2} - \frac{i \theta_2}{16
\pi^2}\right ) \int_{\mathscr{M}} F^{IJ} [{^+A}] \wedge F_{IJ}[{^+A}]
\nonumber\\
&& \mbox{} - \left ( \frac{\theta_1}{8\pi^2} + \frac{i \theta_2}{16
\pi^2}\right ) \int_{\mathscr{M}} F^{IJ} [{^-A}] \wedge F_{IJ}[{^-A}] \, .
\end{eqnarray}
Case I. If $i a_2 = a_1$ and $\theta_2=2 i \theta_1 $, last action acquires
the form
\begin{eqnarray}
S[{^+A}, {^+B}] = 2 a_1 \int_{\mathscr{M}} {^{+} B^{IJ}} \wedge F_{IJ} [{^+A}]
- \frac{\theta_1}{4 \pi^2} \int_{\mathscr{M}} F^{IJ} [{^+A}] \wedge
F_{IJ}[{^+A}] \, ,
\end{eqnarray}
which involves just the self-dual variables.\\
Case II. If $i a_2 = - a_1$ and $\theta_2 =-2i \theta_1$, action
(\ref{previa}) reduces to
\begin{eqnarray}
S[{^-A}, {^-B}] = 2 a_1 \int_{\mathscr{M}} {^{-} B^{IJ}} \wedge F_{IJ} [{^-A}]
- \frac{\theta_1}{4 \pi^2} \int_{\mathscr{M}} F^{IJ} [{^-A}] \wedge
F_{IJ}[{^-A}] \, ,
\end{eqnarray}
which involves just the anti-self-dual variables. Cases I and II involve
either the self-dual or the anti-self-dual connection. In contrast to them,
cases III and IV involve both types of connections:\\
Case III. If $i a_2 = a_1$ and $\theta_2 =-2i \theta_1$, action (\ref{previa})
acquires the form
\begin{eqnarray}
S[{^+A}, {^-A}, {^+B}] & = & 2 a_1 \int_{\mathscr{M}} {^{+} B^{IJ}} \wedge
F_{IJ} [{^+A}] - \frac{\theta_1}{4 \pi^2} \int_{\mathscr{M}} F^{IJ} [{^-A}]
\wedge F_{IJ}[{^-A}]\, ,
\end{eqnarray}
which is the self-dual BF action plus the characteristic of the anti-self-dual
connection.\\
Case IV. If $i a_2 = - a_1 $ and $\theta_2=+2i \theta_1$, action
(\ref{previa}) reduces to
\begin{eqnarray}
S [{^+A},{^-A},{^-B}] = 2 a_1 \int_{\mathscr{M}} {^{-} B^{IJ}} \wedge F_{IJ}
[{^-A}] - \frac{\theta_1}{4 \pi^2} \int_{\mathscr{M}} F^{IJ} [{^+A}] \wedge
F_{IJ}[{^+A}] \, ,
\end{eqnarray}
which involves the anti-self-dual BF theory plus the characteristic based on
the self-dual connection.

\section{Adding quadratic terms in $B^{IJ}$}\label{IV}
Action (\ref{a4}) is not the most general quadratic action that it is possible
to build with the metrics $k_{IJKL}$ and $\varepsilon_{IJKL}$ on the Lie
algebra $so(3,1)$ or $so(4)$, and with the fields $B^{IJ}$ and the curvature
$F^I\,_J [A]$. In fact, it is also possible to consider the action
\begin{eqnarray}\label{a5}
S_4 [A,B] &=& a_1 \int_{\mathscr{M}} B^{IJ} \wedge F_{IJ} [A] + a_2
\int_{\mathscr{M}} {\ast B}^{IJ} \wedge F_{IJ}[A]
\nonumber\\
&& \mbox{} + b_1 \int_{\mathscr{M}} B^{IJ} \wedge B_{IJ} +
b_2 \int_{\mathscr{M}}  {\ast B}^{IJ} \wedge B_{IJ} \nonumber\\
&& \mbox{} + \frac{\theta_1}{8\pi^2} \int_{\mathscr{M}} F^I\,_J [A] \wedge
F^J\,_I [A] + \frac{\theta_2}{16 \pi^2} \int_{\mathscr{M}} {\ast F}^{IJ}[A]
\wedge F_{IJ} [A] \, .
\end{eqnarray}
Thus, there are {\it two} cosmological constants allowed. Their presence, of
course, modifies the space of solutions, which is given by the solutions to
\begin{eqnarray}\label{bf+cc+tt}
{\mathscr E}_{IJ} &=& a_1 F_{IJ} [A] + a_2 {\ast F}_{IJ}[A] + 2 b_1 B_{IJ} + 2
b_2
{\ast B}_{IJ} =0, \nonumber\\
{\cal E}^{IJ} &=& a_1 D B^{IJ} + a_2 D {\ast B}^{IJ} - \frac{\theta_1}{4
\pi^2} D F^{IJ} [A] + \frac{\theta_2}{8\pi^2} D {\ast F}^{IJ} [A] =  0 \, .
\end{eqnarray}
The particular action obtained by setting $a_2 =0$ and $b_2=0$ in equation
(\ref{a5}) has been already reported \cite{4guys}. From the current analysis,
it is clear that the symplectic structure on the space of solutions is simply
the pullback to it of the curl of the symplectic potential given in equation
(\ref{4pot}), taking equations (\ref{bf+cc+tt}) into account.

Action (\ref{a5}) can also  be described by self-dual, anti-self-dual or real
fields following the same lines of section \ref{III}. By decomposing the
fields into their self-dual and anti-self-dual parts, action (\ref{a5})
acquires the form (restricting the analysis to the Lorentzian signature
$\sigma=-1$)
\begin{eqnarray}
S [A,B] &=& ( a_1 + i a_2 ) \int_{\mathscr{M}} {^{+} B^{IJ}} \wedge F_{IJ}
[{^+A}] + ( a_1 - i a_2 ) \int_{\mathscr{M}} {^{-}
B^{IJ}} \wedge F_{IJ} [{^-A}] \nonumber\\
&& \mbox{} + ( b_1 + i b_2 ) \int_{\mathscr{M}} {^{+} B^{IJ}} \wedge
{^+B}_{IJ} + ( b_1 - i b_2 ) \int_{\mathscr{M}} {^{-}
B^{IJ}} \wedge {^-B}_{IJ} \nonumber\\
&& \mbox{} - \left ( \frac{\theta_1}{8\pi^2} - \frac{i \theta_2}{16
\pi^2}\right ) \int_{\mathscr{M}} F^{IJ} [{^+A}] \wedge F_{IJ}[{^+A}]
\nonumber\\
&& \mbox{} - \left ( \frac{\theta_1}{8\pi^2} + \frac{i \theta_2}{16
\pi^2}\right ) \int_{\mathscr{M}} F^{IJ} [{^-A}] \wedge F_{IJ}[{^-A}] \, ,
\end{eqnarray}
from which the various actions involving self-dual and anti-self-dual
variables can be extracted. In the resulting actions, the self-dual and
anti-self-dual variables can be taken as independent variables in the
corresponding action principles. Moreover, it is also possible to consider
real variables choosing properly the parameters involved. We end this section
by rewriting action (\ref{a5}) in the form
\begin{eqnarray}
S [A,B] & = & a_1 \int_{\mathscr{M}} s_{IJKL} B^{IJ} \wedge F^{KL} [A] + b_2
\int_{\mathscr{M}} t_{IJKL} B^{IJ} \wedge B^{KL}
\nonumber\\
& & \mbox{} - \frac{\theta_1}{8 \pi^2} \int_{\mathscr{M}} g_{IJKL} F^{IJ} [A]
\wedge F^{KL} [A] \, ,
\end{eqnarray}
where
\begin{eqnarray}
t_{IJKL} := \frac12 \left ( k_{IJKL} + \frac{b_2}{b_1} \varepsilon_{IJKL}
\right )\, ,
\end{eqnarray}
is a metric on the Lie algebra $so(3,1)$ of $SO(3,1)$. Once again, if
\begin{eqnarray}
\frac{a_2}{a_1} = - \frac{\theta_2}{2 \theta_1} = \frac{b_2}{b_1},
\end{eqnarray}
then there is a single metric on the Lie algebra $so(3,1)$ or $so(4)$.

\section{Concluding remarks}
By using the various metrics defined on the Lie algebra of the internal gauge
group, it is possible to use them to build different action principles which
share the same set of equations of motion but provide different symplectic
geometries on the space of solutions. This is indeed the case for SO(3,1) and
$SO(4)$ four-dimensional BF theories analyzed in this paper where the use of
the Kiling-Cartan metric $\eta_{IJKL}$ and the metric $\varepsilon_{IJKL}$
yields to the symplectic structures ${\bf \Omega}$, ${\bf \Omega}_1$, and
${\bf \Omega}_2$. The inclusion of the second Chern class $c_2 [A]$ and the
Euler class $e[A]$ leads to a relationship between the parameters that appear
in the action which is relevant even classically.

We conclude by making some comments about gravity. If $B^{IJ} = \ast \left (
e^I \wedge e^J \right )$ where $e^I$ is an orthonormal (inverse) tetrad field
is inserted into action (\ref{a3}), Holst's action \cite{Holst}
\begin{eqnarray}\label{palatini}
S[e,A] & = & a_1 \int_{\mathscr{M}} \ast (e^I \wedge e^J ) \wedge F_{IJ} [A] +
\sigma a_2 \int_{\mathscr{M}} e^I \wedge e^J \wedge F_{IJ}[A]\, ,
\end{eqnarray}
is obtained while if this substitution is done in action (\ref{a4}), action
(\ref{palatini}) complemented with topological terms is obtained \cite{Mon}.
On the other hand, if they are inserted into the action (\ref{a5}), a
cosmological term is added to the action of \cite{Mon}
\begin{eqnarray}\label{xxx}
S[e,A] &=&  a_1 \int_{\mathscr{M}} \ast (e^I \wedge e^J ) \wedge F_{IJ} [A]
+ \sigma a_2 \int_{\mathscr{M}} e^I \wedge e^J \wedge F_{IJ}[A] \nonumber\\
&& \mbox{} + \frac{\sigma b_2}{2} \int_{\mathscr{M}} \varepsilon_{IJKL} e^I
\wedge e^J \wedge
e^K \wedge e^L \nonumber\\
&& \mbox{} + \frac{\theta_1}{8\pi^2} \int_{\mathscr{M}} F^I\,_J [A] \wedge
F^J\,_I [A] + \frac{\theta_2}{16 \pi^2} \int_{\mathscr{M}} {\ast F}^{IJ}[A]
\wedge F_{IJ} [A] \, .
\end{eqnarray}
In particular, note that if $\frac{1}{\beta} := \frac{a_2}{a_1} = - \sigma
\frac{2 \theta_1 }{\theta_2}$, where $\beta$ is the Barbero-Immirzi parameter
then one also has a single metric on the Lie algebra $so(3,1)$ or $so(4)$
which allows it to introduce a particular set of canonical coordinates-the
analogue of the canonical pairs (\ref{canc}) for BF theory. $\beta$ is related
to the other parameters (especially interesting is the case where $\beta$ is a
real number in the case of $SO(3,1)$ or $\beta \neq \pm 1$ in the case of
$SO(4)$). The condition $B^{IJ}=\ast (e^I \wedge e^J)$ can be added to action
(\ref{a5}) via $\lambda_{IJ} \wedge \left ( B^{IJ} - \ast
 (e^I \wedge e^J) \right )$ where $\lambda_{IJ}=-\lambda_{JI}$ are Lagrange
 multipliers 2-forms.

\section*{Acknowledgements}
The author thanks the comments of A. Corichi, M. Mondrag\'on, A. Perez, C.
Rovelli, and J.A. Zapata at the poster session of the meeting LOOPS '05 held
at the Max-Planck-Institute for Gravitational Physics, Golm, Berlin, Germany,
October 2005. The author also thanks R. Capovilla, G.F. Torres del Castillo,
and J.D. Vergara for very fruitful discussions. This work was supported in
part by CONACyT grant no. SEP-2003-C02-43939.

\section*{Appendix}\label{notes}
The infinitesimal generators of $SO(4)$ or $SO(3,1)$ are the boosts $b_i$ and
the rotations $r_i$ which in the 4-dimensional representation are given by
\begin{eqnarray}
b_1 = \left (
\begin{array}{rrrr}
0 & 1 & 0 & 0 \\
-\sigma & 0 & 0 & 0 \\
0 & 0 & 0 & 0 \\
0 & 0 & 0 & 0
\end{array}
\right ), b_2 = \left (
\begin{array}{rrrr}
0 & 0 & 1 & 0 \\
0 & 0 & 0 & 0 \\
-\sigma & 0 & 0 & 0 \\
0 & 0 & 0 & 0
\end{array}
\right ), b_3 = \left (
\begin{array}{rrrr}
0 & 0 & 0 & 1 \\
0 & 0 & 0 & 0 \\
0 & 0 & 0 & 0 \\
-\sigma & 0 & 0 & 0
\end{array}
\right ),
\end{eqnarray}
and
\begin{eqnarray}
r_1 = \left (
\begin{array}{rrrr}
0 & 0 & 0 & 0 \\
0 & 0 & 0 & 0 \\
0 & 0 & 0 & -1 \\
0 & 0 & 1 & 0
\end{array}
\right ), r_2 = \left (
\begin{array}{rrrr}
0 & 0 & 0 & 0 \\
0 & 0 & 0 & 1 \\
0 & 0 & 0 & 0 \\
0 & -1 & 0 & 0
\end{array}
\right ), r_3 = \left (
\begin{array}{rrrr}
0 & 0 & 0 & 0 \\
0 & 0 & -1 & 0 \\
0 & 1 & 0 & 0 \\
0 & 0 & 0 & 0
\end{array}
\right ),
\end{eqnarray}
which satisfy
\begin{eqnarray}
\left [ r_i , r_j \right ] = \varepsilon_{ij}\,^k r_k, \quad \left [ b_i , b_j
\right ] = \sigma \varepsilon_{ij}\,^k r_k, \quad \left [ r_i , b_j \right ] =
\varepsilon_{ij}\,^k b_k,
\end{eqnarray}
$i,j,k=1,2,3$, with $(\eta_{IJ})=\mbox{diag}(\sigma,1,1,1)$ with $\sigma=1$
for the Euclidean and $\sigma=-1$ for the Minkowskian internal metrics
$\eta_{IJ}$. Equivalently,
\begin{eqnarray}
X_{0i}= b_i, \quad X_{ij}= - \varepsilon_{ij}\,^k r_k,
\end{eqnarray}
with
\begin{eqnarray}
[ X_{IJ} , X_{KL} ] = \eta_{JK} X_{IL} - \eta_{IK} X_{JL} - \eta_{JL} X_{IK} +
\eta_{IL} X_{JK}.
\end{eqnarray}

In particular, the connection 1-form $A^I\,_J$ has the matrix representation
\begin{eqnarray}
\left ( A^I\,_J \right ) = \left (
\begin{array}{cccc}
0 & A^0\,_1 & A^0\,_2 & A^0\,_3 \\
-\sigma A^0\,_1 & 0 & A^1\,_2 & A^1\,_3 \\
-\sigma A^0\,_2 & - A^1\,_2 & 0 & A^2\,_3 \\
-\sigma A^0\,_3 & - A^1\,_3 & - A^2\,_3 & 0
\end{array}
\right ) = \Gamma^i r_i + A^{0i} b_i,
\end{eqnarray}
with $\Gamma^i= -\frac12 \epsilon^i\,_{jk} A^{jk}$, with similar conventions
for $B^I\,_J$ and $F^I\,_J [A]$.

The components, $\varepsilon_{IJKL}$, of the metric tensor $\varepsilon$ in
the basis $X_{IJ}$, are given by $\varepsilon \left ( X_{IJ}, X_{KL} \right )=
\varepsilon_{IJKL}$ with $\varepsilon_{0123}=\epsilon$. In matrix form
\begin{eqnarray}
\left ( \varepsilon_{IJKL} \right )= \epsilon \left (
\begin{array}{cccccc}
0 & 0 & 0 & 1 & 0 & 0 \\
0 & 0 & 0 & 0 & 1 & 0 \\
0 & 0 & 0 & 0 & 0 & 1 \\
1 & 0 & 0 & 0 & 0 & 0 \\
0 & 1 & 0 & 0 & 0 & 0 \\
0 & 0 & 1 & 0 & 0 & 0
\end{array}
\right ),
\end{eqnarray}
with $\epsilon=1$ or $\epsilon=-1$ depending on the orientation of the
internal volume element. The couples $IJ$ and $KL$ take on the values $01$,
$02$, $03$, $23$, $31$, and $12$. On the other hand, for the Killing-Cartan
metric $k (X_{IJ}, X_{KL})=k_{IJKL}=\eta_{IK}\eta_{JL}-\eta_{JK}\eta_{IL}$ one
has
\begin{eqnarray}
\left ( k_{IJKL} \right ) = \left (
\begin{array}{cccccc}
\sigma & 0 & 0 & 0 & 0 & 0 \\
0 & \sigma & 0 & 0 & 0 & 0 \\
0 & 0 & \sigma & 0 & 0 & 0 \\
0 & 0 & 0 & 1 & 0 & 0 \\
0 & 0 & 0 & 0 & 1 & 0 \\
0 & 0 & 0 & 0 & 0 & 1
\end{array}
\right ).
\end{eqnarray}
The determinant of the metric $k_{IJKL} + \alpha \varepsilon_{IJKL}$ is
$\det\left( k_{IJKL} + \alpha \varepsilon_{IJKL} \right )= - \alpha^6 + 3
\sigma \alpha^4 - 3 \alpha^2 + \sigma$ (i.e., it is independent of the
orientation $\epsilon$). It vanishes if and only if $\alpha^2=\sigma$ which
corresponds to the self-dual and anti-self-dual cases.


\end{document}